\documentclass[compsoc,conference,a4paper,10pt]{IEEEtran}
\IEEEoverridecommandlockouts
\usepackage{cite}
\usepackage{amsmath,amssymb,amsfonts}
\let\phi\varphi 
\usepackage{booktabs} 
\usepackage{multirow} 
\usepackage{pifont} 
\usepackage{algorithmic}
\usepackage{graphicx}
\usepackage{textcomp}
\usepackage{bmpsize}
\usepackage{xcolor}
\usepackage{lipsum}
\usepackage[colorlinks=true,urlcolor=black]{hyperref}
\def\BibTeX{{\rm B\kern-.05em{\sc i\kern-.025em b}\kern-.08em
    T\kern-.1667em\lower.7ex\hbox{E}\kern-.125emX}}
\usepackage{cleveref} 
\begin{document}

\makeatletter
\newcommand{\linebreakand}{%
  \end{@IEEEauthorhalign}
  \hfill\mbox{}\par
  \mbox{}\hfill\begin{@IEEEauthorhalign}
}
\makeatother

\title{Deep Down the Rabbit Hole: \\
On References in Networks of Decoy Elements}

\author{
\IEEEauthorblockN{Daniel Reti,
Daniel Fraunholz,
Janis Zemitis,
Daniel Schneider and
Hans Dieter Schotten} \\
\IEEEauthorblockA{Intelligent Networks Research Group\\
German Research Center for Artificial Intelligence\\
67655 Kaiserslautern, Germany\\
Email: \{firstname\}.\{lastname\}@dfki.de}}

\maketitle

\begin{abstract}
Deception technology has proven to be a sound approach against threats to information systems. Aside from well-established honeypots, decoy elements, also known as honeytokens, are an excellent method to address various types of threats. Decoy elements are causing distraction and uncertainty to an attacker and help detecting malicious activity. Deception is meant to be complementing firewalls and intrusion detection systems. Particularly insider threats may be mitigated with deception methods. While current approaches consider the use of multiple decoy elements as well as context-sensitivity, they do not sufficiently describe a relationship between individual elements. In this work, inter-referencing decoy elements are introduced as a plausible extension to existing deception frameworks, leading attackers along a path of decoy elements. A theoretical foundation is introduced, as well as a stochastic model and a reference implementation. It was found that the proposed system is suitable to enhance current decoy frameworks by adding a further dimension of inter-connectivity and therefore improve intrusion detection and prevention.
\end{abstract}


\section{Introduction}\label{sec.introduction}
The widely adopted perimeter-based and signature-based defense strategies in network security have a passive nature of preventing and detecting unauthorized resource access. While network firewalls and Intrusion Detection Systems (IDS) are theoretically quite capable of effectively preventing a compromise, in reality, sophisticated and persistent attackers often manage to find previously unconsidered attack vectors. These might be, but are not limited to social engineering, weak credentials or the misconfiguration of defense measures. Strongly depending on the network's security plan, an attacker who manages to gain network access is hard to detain and will likely cause disastrous consequences. A similar risk is posed by a malicious insider with legitimate network access and the capabilities seeing and accessing other hosts in the network.

While many defense-in-depth strategies are capable of mitigating insider threats, this work applies the strategy of deception technology, which is most commonly known for honeypots but also encompasses other types of fake entities. In contrast to conventional security measures, deception has a more active nature regarding the interaction with attackers, aiming to distract an attacker away from real targets towards decoy targets. Decoy elements, also referred to as honeytokens or honeytraps, proposed in previous research include leaking fake accounts, credentials, documents, executables or database entries \cite{Fraunholz.2018demystifying}.

The novel approach of this work is to let different types of decoy elements have references to other decoy elements with the goal of forming a well-defined network of inter-referencing decoy elements, leading the attacker along a specific path towards a honeypot or a less critical network segment, while exhausting the attackers time and resources. Such a network of referencing decoy elements is also useful to detect perimeter breaches and trigger adequate containment measures and thus reduce the probability of successful privilege escalation or lateral movement. 

In recent research, frameworks for the deployment of such elements were proposed. These frameworks particularly cope with the challenge of context-sensitivity for the deployment of decoy elements \cite{Fraunholz.2018demystifying}. However, to the best of the authors’ knowledge, no previous research was conducted on networks that consist of decoy elements referencing further decoy elements. In this work the authors provide the theoretical foundation for a generation framework for inter-referencing decoy element networks as well as a qualitative evaluation of the proposed framework and reference-implementation.
This work is structured as follows: In section 2, the theoretical background of deception and decoys as well as the state of the art is introduced. After that, the proposed system of graph computation and transition probabilities is described in section 3. Subsequently, the reference implementation is presented in section 4. In section 5, an evaluation of the proposed system is given. A conclusion is given in section 6.

\section{Background}\label{sec.background}
In this section the fundamentals of deception technology are presented and for a proper problem statement and evaluation the applied
attacker model is introduced.

\subsection{Deception Technology}\label{subsec.deception}

Deception offers a diverse set of methods for information security. Although deception is typically targeting human adversaries, software can be deceived as well. Web application fuzzers as an example rely strongly on feedback in form of status codes, version information and banners which can easily be manipulated without a degradation in usability \cite{Fraunholz.2018demystifying, Fraunholz.2018defending, Nawrocki.2016}. A typical objective is to increase the interaction between an intruder and a deception system and thus exhaust resources, such as time or computational power, and expose tactics, techniques and procedures (TTPs). The most common taxonomy for deception, as proposed by \cite{Bell.1991}, distinguishes between simulation and dissimulation. Simulation-based deception is defined by presenting something that does not exist, utilizing the strategies of mimicking, inventing and decoying.  A well-known application of simulation-based deception is the use of honeypots and honeynets. The decoy elements used in this work are all simulation-based deception methods. Dissimulation-based methods aim to conceal the truth by masking, repacking and dazzling. Commonly known dissimulation-based deception techniques are obfuscation and moving target defense.

\subsection{Attacker Model and Problem Statement}\label{subsec.attacker}
This work is approaching an attacker who has gained access to the network in which the proposed system is operating. Typical scenarios are a malicious insider or a pivoting attacker. Although the system could be configured to be applicable for a perimeter system, this is not the focus of this work. The attacker performs reconnaissance and enumerates hosts and services on the network before and attack is chosen, therefore attack vectors of an attacker in such a position are strongly dependent on the system's information. Gathered system information are typically used to identify user accounts, other hosts in the network, unprotected services or further potential attack vectors. Therefore, the proposed system is aware of common information sources of interest for attackers, where it presents false information, increasing the complexity for an attacker to identify relevant information and furthermore guiding an attacker on wrong tracks of misleading information.

\subsection{Decoy Elements}\label{subsec.decoy}
While the concept of decoy elements is nothing new, the term honeytoken as a reference to honeypots was only coined in 2003 by DeBarros \cite{DeBarros.2007}, since then it was implemented in various shapes and environments. The usage of decoy elements is limited rather by imagination than by technical restrictions. Any type of information may be applicable as a decoy element. Previous studies present an overview of common decoy elements \cite{Fraunholz.2018demystifying, Virvillis.2014}. In this work, a set $\Omega$ of 10 decoy element types is considered for further analysis. While no novel decoy types are introduced in this work, a classification is proposed in order to describe properties that are introduced with inter connectivity of decoy elements. First, decoy elements may be accessed from different kinds of proximity. A suitable classification was adapted from the Common Vulnerability Scoring System (CVSS) \cite{CVSS.2018}. The groups are: Physical (A1), local (A2), adjacent (A3) and network (A4). Second, each type has a set of features describing essential properties during deployment and operation.

\begin{itemize}
	\item[(a)] Self-referenceable (P1): Property of the decoy element to be able to reference to a decoy element of its own type. This feature is important for the generation of the reference network. If this property is not given, the decoy type cannot be subsequent to a decoy of the same type. An example for non-self-referenceable is an ARP-cache entry. Such entries can only reference to IP- and hardware addresses but not to further cache entries.
	\item[(b)] Observable (P2): Several decoy types cannot be directly observed during operation. This property depends on the given scenario, as a particular type can be observable in one scenario but not in another. An example is a DB-entry, which may be non-observable if a MySQL-client with encrypted communication is used, but is observable if the communication is unencrypted, as the queries can be monitored with tools such as \textit{wireshark} or \textit{tcpdump}.
	\item[(c)] Runtime revisions (P3): An important property in dynamic deployment environment is the ability of a decoy element to be revised during runtime. For this property, plausibility
must be taken into account, since from a technical perspective most decoy types may be revised during operation. However, for types with a dynamic nature, such as the caches, this property is more likely and thus more believable.
	\item[(d)] Post-deployment revisions (P4): This property is related to the previous one, but has the requirement of being able to be revised after the deployment and operation, but not during operation.
	\item[(e)] Privileged (P5): Several decoy types are only accessible by privileged users. An example of such a decoy type is the ARP-cache and the shadow file \cite{Zohar}.
	\item[(f)] Multi-domain (P6): A decoy element can be configured to be accessed by either one or multiple entities, depending on the reading permissions. In the presented work, a domain is implemented as a user or group that is able to access a particular resource. This property describes the ability to be accessible by more than one particular user without administrative privileges. Typically user logs are only readable by the user itself or administrators and would therefore not fulfill this property.
	\item[(g)] Multi-instance (P7): This property describes whether a decoy element type might be deployed more then once on a system. An operating system usually only has one host file, thus the host file as a decoy element does not have this property, while one host file may contain multiple entries as references to other decoy elements. 
\end{itemize}

Third, each decoy type has a particular set of decoy elements it
can reference to. For example, a document may reference to most
other decoy types, whereas a host file is only able to reference to
domains and IP addresses. This kind of limitation must be taken
into account when generating the decoy networks. An exemplary, incomplete set of decoy
types considered in this work is given in \cref{tab.decoys} and described in the following. The selection of decoy element types is based on information the authors expect to be of interest for an attacker.

\begin{table*}[h!t] 
	\renewcommand{\arraystretch}{1.3}
	\caption{Properties of potential decoy elements}
	\label{tab.decoys}
	\centering
	\begin{tabular}{l l l l l c c c c c c c}
		\toprule
		\multicolumn{2}{c}{\textbf{Meta information}} & \phantom{a} & \multicolumn{6}{c}{\textbf{Properties}}\\
		Name & Abbrev. & & Access type & Outbound connections & P1 & P2 & P3 & P4 & P5 & P6 & P7\\
		\cmidrule{1-2} \cmidrule{4-12}\\
		Comment \cite{Fraunholz.2018defending} & $D_{1}$ & & A2, A3, A4 & $D_{1}$-$D_{10}$ & \ding{51} & \ding{51} & \ding{55} & \ding{55} & \ding{55} & \ding{51} & \ding{51}\\
		DB-entry \cite{Bercovitch.2011, Fraunholz.2018defending, Hoglund.2011, Krawetz.2004, Virvillis.2014} & $D_{2}$ & & A2, A3, A4 & $D_{2}$-$D_{4}$, $D_{6}$, $D_{8}$-$D_{10}$ & \ding{51} & \ding{51} & \ding{55} & $\bigcirc$ & \ding{55} & \ding{51} & \ding{51}\\
		ARP-cache  \cite{Zohar} & $D_{3}$ & & A2 & $D_{8}$, $D_{10}$ & \ding{55} & \ding{51} & \ding{51} & \ding{51} & \ding{51} & \ding{55} & \ding{55}\\
		Account \cite{Almeshekah.2015, Fraunholz.2018defending, Juels.2013} & $D_{4}$ & & A2, A3, A4 & $D_{1}$-$D_{10}$ & \ding{51} & \ding{51} & \ding{51} & \ding{55} & \ding{55} & \ding{51} & $\bigcirc$ \\
		History \cite{Zohar} & $D_5$ & & A2 & $D_1$-$D_{10}$ & \ding{51} & $\bigcirc$ & \ding{51} & \ding{51} & \ding{55} & \ding{55} & $\bigcirc$ \\
		Document \cite{Fraunholz.2018defending, Lazarov.2016, Virvillis.2014, Zohar} & $D_6$ & & A2, A3, A4 & $D_1$-$D_{10}$ & \ding{51} & \ding{51} & \ding{51} & \ding{55} & \ding{55} & \ding{51} & \ding{51} \\
		File-Metadata \cite{Rowe.2007} & $D_7$ & & A2, A3, A4 & $D_2$, $D_4$, $D_6$, $D_8$ & \ding{55} & \ding{51} & \ding{51} & \ding{55} & \ding{55} & \ding{51} & $\bigcirc$ \\
		URL/Link \cite{Fraunholz.2018defending, Zohar} & $D_8$ & & A3, A4 & $D_4$, $D_6$, $D_{10}$ & \ding{51} & \ding{51} & \ding{51} & \ding{55} & \ding{55} & \ding{51} & \ding{51} \\
		Host file & $D_9$ & & A2 & $D_4$, $D_8$, $D_{10}$ & \ding{51} & \ding{51} & \ding{55} & \ding{55} & \ding{51} & $\bigcirc$ & $\bigcirc$ \\
		Port \cite{Fraunholz.2018defending, Zohar} & $D_{10}$ & & A3, A4 & $D_4$, $D_8$, $D_{10}$ & \ding{51} & \ding{51} & \ding{51} & \ding{55} & \ding{55} & \ding{51} & \ding{51} \\
		\bottomrule
	\end{tabular}
\end{table*}

\begin{itemize}
	\item[(a)] Comment ($D_1$): Comments may contain any kind of information and are commonly found in source code or configuration
    files. In contrast to documents-type decoys, comment-type decoys are not part of the file's content, but rather a remark to the content and thus not interfering with regular usage or functionality. This type of decoy can reference most other decoy types, since comments are flexible in their content. When comments are deployed into existing content, context-sensitivity is required. This can be achieved by using either generic comments or content analysis, e.g. with NLP methods \cite{Whitham.2017}.
	\item[(b)] DB-entry ($D_2$): Entries in a DB are similarly flexible as comments and require the same considerations regarding their deployment. Furthermore, it must be taken into account that remote access to a DB is rather usual. It is also important to consider the observability. Monitoring access to a DB file is less complex than access to a particular table, which itself is less complex than monitoring access to a particular decoy entry. The later ones are not observable based on file access, instead log monitoring or network sniffing may be applied to monitor access \cite{Krawetz.2004}.
	\item[(c)] ARP cache ($D_3$): In contrast to the former decoy types, the ARP cache has a limited set of information it can provide. The most conclusive information from an attacker’s perspective may be the IP and MAC address contained in an entry. This decoy type can be used to reference network addresses, which can be located on the local or a remote system. As stated in \cref{tab.decoys}, administrative privileges are required to directly access information in the ARP cache. This has the disadvantage of a decreased probability of detecting unprivileged attackers, but also provides the insight that usage of the deployed information must be the consequence of a full system compromise. However, monitoring the ARP cache based on file access is not possible, as the cache file is frequently accessed for legitimate reasons.
	\item[(d)] Account ($D_4$): Accounts can be used as versatile decoy elements. They can be referenced in various contexts and often do not require more information than a user name. In cases where authentication software can be manipulated (e.g. PAM in Unix), authentication attempts into accounts may be simple to monitor. Furthermore, they frequently provide complete sets of files, directories or even environments to the logged-in entity, thus enabling a broad range of further deception. As account and password management is still a ongoing challenge in information security, references to accounts, including credentials, may be not susceptible.
	\item[(e)] History ($D_5$): Several types of software create and store history information. In this work, history files used for optimized user experience (browser history) and convenience (shell history) are taken into account. History files used for traceability and accountability such as log files are not considered, even though they are applicable as decoy elements.
	\item[(f)] Document contents ($D_6$): Documents are the best--known types of decoy elements. Document contents can contain any type of information. They are observable by means of file-access monitoring, and more flexible than comments and DB-entries, but have the downside of interfering with usage and might be mistakenly found by legitimate users. Typically, the most effort for this decoy type is devoted to contextual adaptation \cite{Whitham.2017}, which is not part of this work.
	\item[(g)] File metadata ($D_7$): Information that is not part of the content itself is frequently of sensitive nature as well. In this work, no aggregation or inference of sensitive information is considered. However, file names often leak information about file content or even about further files or directory structures. Further metadata, such as time stamps, file owner or groups may reveal information, e.g. the existence of user accounts or working hours of particular users.
	\item[(h)] Uniform Resource Locator ($D_8$): Given the definition in RFC 3986 section 3, a Uniform Resource Locator (URL) follows the syntax:

	\begin{center}
		{\small \texttt{URL = scheme : [//authority]path[?query][\#f ragment],}} \\ 		
		where an authority consists of the following parts: \\ 	
		{\small \texttt{authority = [userinfo@]host[:port.}}
	\end{center}	
	
	This syntax can be exploited to deploy decoy elements of various types within a URL. Several components are directly indicating a suitable decoy type such as: Path, userinfo, host and port. However, the query and fragment component can be used to reference any other decoy type. A query component typically consists of a parameter and value that can be chosen arbitrarily, while the fragment component refers to a section within a resource, which can also be chosen arbitrarily.
	\item[(i)] Host file ($D_9$): Similar to the ARP cache, a host file is limited in the kind on information it provides and the ability to monitor access. The major difference to the ARP cache is that instead of a MAC address, a domain is stored. Host files may contain comments, but they are not considered in this work.
	\item[(j)] Port ($D_{10}$): Most TCP ports are associated with particular services. However, information about a service may also be obtained by connecting to an open port, where services may transfer a protocol specific message, containing protocol or service name and version information, followed by a message. Plausible messages are redirects to other hosts or ports as well as error messages referring to a contact which could be an email address.
\end{itemize}
An extensive overview of further decoy types is given by \cite{Fraunholz.2018demystifying} and \cite{Virvillis.2014}. The terminology is discussed by \cite{Pouget.2003}.

\section{The Rabbit Hole}\label{sec.rabbit}
In this section, the framework for the generation and operation of inter-referencing decoy networks is introduced. First, the computation of the network topology is discussed. Then, the probabilistic model to determine the network population and the references is defined. The introduced framework is also able to generate references at runtime, which is discussed at the end of this section. 

\subsection{Computation of Directed Graphs}\label{subsec.computation}
A reference has a directional property. Therefore, the network generated has to be directional at initialization as well. The network is represented as a graph, where a decoy element is a node and a reference is a directed edge. The directed graph is generated by a graphical model. As there are several models and assigned properties, the requirements of the network must be defined in advance, in order to select a model accordingly.

\begin{itemize}
	\item[(a)] Directed edges: As previously discussed, a reference is information at a particular location, pointing to another particular location. From this definition, it can be concluded that edges need to be directional.
	\item[(b)] Reciprocal edges: It is not required that the referred location contains information referring back to the previous location. Like other references, reciprocal references increase the probability to lure a trespasser into the decoy network.
	\item[(c)] Acyclic property: This property may affect subsequent processing steps and needs to be considered, e.g. when traversing the graph with a depth-first search, there is no restriction from a information security perspective. In particular, loops with several nodes may mislead a trespasser into more interaction.
	\item[(d)] Connectivity: It is not required that there is a particular k-connectivity, neither for nodes nor for edges. However, the improvement proposed in this work is the concept of references, which are supposed to drag a trespasser into the decoy network and keep the trespasser engaged with reconnaissance activities. Therefore, in this work, connectivity is assumed to be required for the generated graph.
	\item[(e)] Adjacency: The number of in- and outbound references for each decoy is dependent on the decoy type itself. However, as discussed in detail later on, a trespasser may not be able to follow a particular reference for lack of knowledge or permission. It is therefore considered more effective to ensure a equal distribution of in- and outbound references.
\end{itemize}

Given these identified restrictions, there is still flexibility for the
generation of graphs. \Cref{fig.valid} gives four examples of valid graphs
from four different graphical models.

\begin{figure}
	\caption{Examples of valid graphs, based on the networkx implementation of the named models, each with 10 nodes, Erd{\H o}s-R{\' e}nyi with $p = 0.15$, random k-out with $k = 1$.}
	\label{fig.valid}
	\includegraphics[scale=1]{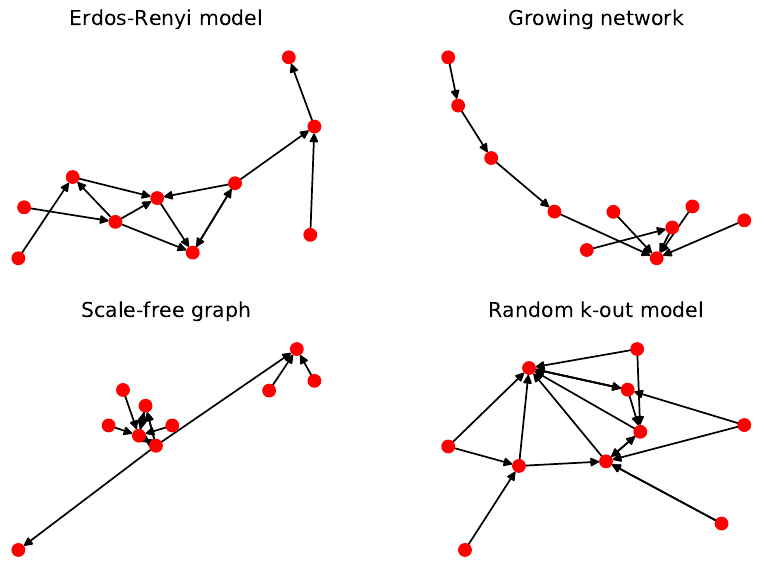}
\end{figure}

As it can be seen, the selection of a graph model remains subject to the implementation. For the reference implementation, the \textit{Erd{\H o}s-R{\'e}nyi} model in the $G(n, p)$ variant \cite{ErdHos.1960} is used to generate random graphs, satisfying the discussed properties. However, there are further models such as \textit{Barab{\'a}si-Albert} and \textit{Watts-Strogatz} that may be used instead. Different models can be randomly selected during initialization or mixed. Further evaluation on the correlation of network generation models and deception success is required. As discussed before, connected graphs are preferred to unconnected graphs. This can be achieved, for example, by selecting only the largest subgraph, either in terms of edges or nodes. Alternatively, the subgraphs may be connected.

\subsection{Probabilistic Model for Decoy Types}\label{subsec.probabilistic}
A time-homogeneous discrete-time \textit{Markov} chain (DTMC) was chosen to represent the model for the generation of decoy type
sequences. This model determines subsequent decoy types to a given decoy type. The probability $p_{ij}$ of decoy type $j$ to be deployed subsequently to decoy type $i$ is defined in \cref{eq.1}.

\begin{equation}\label{eq.1}
	p_{ij} = Pr(X_{t-1} = i|X_t = j)\ \text{for } i,j \in \Omega 
\end{equation}

$p_{ij}$ represents the forward transition probability within the model. However, due to the directed characteristic of the generated graphs, it may be necessary to traverse an edge backwards, in order to identify precedent referencing nodes. The probability of a backward transition is given in \cref{eq.2}.

\begin{equation}\label{eq.2}
	\pi_ip_{ij} = \pi_jp_{ji}
\end{equation}

The necessary and sufficient condition for \cref{eq.2} is \textit{Kolmogorov's} criterion \cite{Kelly.1981} as given in \cref{eq.3}.

\begin{equation}\label{eq.3}
	p_{i_1 i_2}p_{i_2 i_3}...p_{i_{n-1} i_n}p_{i_ni_1} = p_{ji} \forall i_1, i_2 ... i_n \in \Omega
\end{equation}

\textit{Kolmogorov's} criterion requires that the product of probabilities on each closed loop must be equal in both directions. If the criterion is satisfied, $P^* = P$. If it is not reversible, $P^*$ can be computed from $P$ by employing Bayes law as given in \cref{eq.4}.

\begin{equation}\label{eq.4}
	p^*_{ij} = Pr(X_{t-1} = j|X_t = i) \dfrac{P(X_t = E_j)}{P(X_{t-1} = E_i)}
\end{equation}

$P(E)$ is the stationary distribution $\phi^T$ . This distribution is uniform if $P$ is symmetric. It can be determined as given in \cref{eq.5} and \cref{eq.6}.

\begin{equation}\label{eq.5}
	\lim_{N \to \infty} p^N = 
	\begin{pmatrix}
		\phi_1 & \phi_2 & ... & \phi_\Omega \\
		\phi_1 & \phi_2 & ... & \phi_\Omega \\
		\vdots & \vdots & & \vdots \\
		\phi_1 & \phi_2 & ... & \phi_\Omega \\
	\end{pmatrix}
\end{equation}

\begin{equation}\label{eq.6}
	\phi^T = (\phi_1, \phi_2, ..., \phi_\Omega)
\end{equation}

$\phi^T$ is also used as probabilities for the decoy type of the entry node. In doing so, it is ensured to start in the stationary phase of the DTMC. It additionally provides the expected distribution of decoy elements of each type, it can therefore be used to analyze and tweak the distribution of decoy types while configuring the model.

\subsection{Computation Efficient Probability}\label{subsec.efficient}
During the operational phase, the probability for subsequent decoy elements can be adapted to the observed activity. As attackers are more likely to follow references that they have knowledge of and that appear useful, the system can be optimized by increasing the likelihood for transitions that are observed, while decreasing the likelihood for transitions that are ignored. This learning model assumes that the behavior of multiple attackers is not completely unrelated. Therefore, if it were completely unrelated, an adaption to the behavior would not be possible. There are two requirements to be met when designing the learning functionality:

\begin{itemize}
	\item[(a)] The sum of each row in $P$ must be $1$. Therefore an increase in a particular transition probability must result in a decrease of the same magnitude. The decrease can be split between multiple transition probabilities.
	\item[(b)] $p_{ij}$ must be ensured to be in the range $0 \leq p_{ij} \leq 1$.
\end{itemize}

Optionally, the probability adaption $\Delta p_{ij}$ can be dynamically adjusted to the current transition probability $p_{ij}$. This property increases the resiliency against overfitting. Following the design decision not to differentiate between the set of actually available outbound references and the potential set of outbound references as given in \cref{tab.decoys}, a computational efficient learning method is defined in \cref{eq.7} and \cref{eq.8}.

\begin{equation}\label{eq.7}
	p^{t+1}_F = p^{t}_F + \eta (1-p^t_F)
\end{equation}

\begin{equation}\label{eq.8}
	p^{t+1}_A = p^t_A(1-\eta )
\end{equation}

$p^{t}_F$ is the transition probability of the reference followed at the time $t$ and $p^t_A$ the transition probability of references that are not followed at the time $t$. $\eta$ is the learning rate, used to parametrize the impact observations have on the transition probability. In this learning method, a uniformly distributed split of the aforementioned decrease between the ignored references is implemented. \par 
The fulfillment of the first requirement can be proven as shown in \cref{eq.9}, stating that positive deltas on the left side are always the same as the sum of all negative deltas on the right side for all numbers of ignored references.

\begin{equation}\label{eq.9}
	\eta (1-p^t_F) = \sum_\Omega \eta p^t_A \forall|\Omega|\in \mathbb{N}
\end{equation}

To prove the second requirement the relevant limits are determined as shown in \cref{eq.10}.

\begin{equation}\label{eq.10}
	\Delta p_{ij} = 
	\begin{cases}
		\eta(1-p_{ij}) & \text{if followed} \\
		-\eta p_{ij} & \text{if ignored}
	\end{cases}
\end{equation}

There are two identified limits that can be used to prove the fulfillment of the second requirement as given in \cref{eq.11} and \cref{eq.12}. 

\begin{equation}\label{eq.11}
	\lim_{p_{ij} \to 1} \eta(1-p_{ij}) = 0
\end{equation}

\begin{equation}\label{eq.12}
	\lim_{p_{ij} \to 0} -\eta p_{ij} = 0
\end{equation}

In \cref{fig.learning}, an example is visualized for the learning mode. For the sake of clarity, a set of only two potential outbound references is considered. The example shows the transition probabilities and their deltas in respect to the observations made.

\begin{figure*}
	\caption{Example visualization for learning mode, two possible outbound references (orange, blue), reference probability (solid), delta probability (dashed), same reference (left), uniform reference choice (right, down), learning rate $\eta = 0.3$ (up), learning rate $\eta = 0.01$ (down)}
	\label{fig.learning}
	\includegraphics[scale=1]{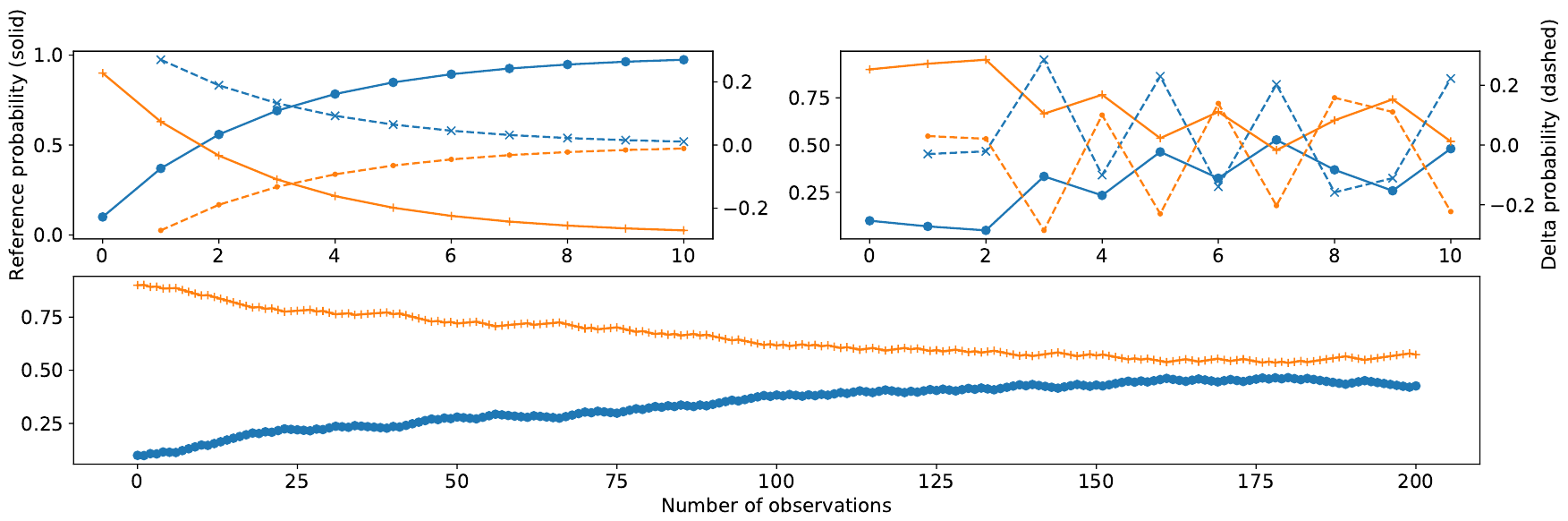}
\end{figure*}

On the top, the probabilities for the references are 0.85 (orange) and 0.15 (blue) with a learning rate $\eta$ of 0.3. On the left side, the probability to observe activity on the blue reference is set to 1, while it is 0 for the orange reference. As proven before, the probabilities asymptotically approach the limit values derived from the requirements. Additionally, the satisfaction of the optional third requirement is visualized by the dashed lines. If the probabilities are more close to the limit values, the changes are decreasing dynamically. On the top right side, the initialization values for the probabilities are equal to the previous example as well as the learning rate. However, in this example, the probability to observe activity on the blue reference is set to 0.5, while it is also 0.5 for the orange reference. As it can be seen, the probabilities do not converge. This is because a learning rate of 0.3 is too high for the number of observations. The learning rate must be adapted to the expected number of observations. In the example on the bottom, a learning rate of 0.01 was chosen, while the initialization values for the probabilities are equal to the previous examples. In order to illustrate the effect of low learning rates, the number of observation was increased to 200. The probability to observe activity on the blue reference is set to 0.5, while it is also 0.5 for the orange reference. As it can be seen, the reference probabilities are approaching the observation probabilities. However, learning rates in this magnitude require a large number of observations, which cannot be expected to occur within the security perimeter.

\section{Implementation}\label{sec.implementation}
The reference implementation is developed for \textit{Unix}-based operating systems. Therefore, some decoy types may not be applicable for other environments. Even though the program is written in \textit{Python 3.5} which is platform independent, operating in other environments may require to employ an alternative solution for file access monitoring and decoy deployment. Several monitoring solutions for \textit{MS-Windows} exist, such as \textit{FileSystemWatcher}, \textit{FindFirstChangeNotification} or \textit{inotify-win}, an \textit{inotify} port to \textit{MS-Windows}. \textit{fsevents} may be applicable for \textit{OSX} environments. \textit{SQLlite} is used to store and manage information about the deployed decoys. \Cref{fig.flow} gives an overview of the program flow.

\begin{figure}
	\caption{Overview of the program flow with focus on the supported operational modes and their differences. Chart is created in compliance with \textit{SDL}.}
	\label{fig.flow}
	\includegraphics[scale=1]{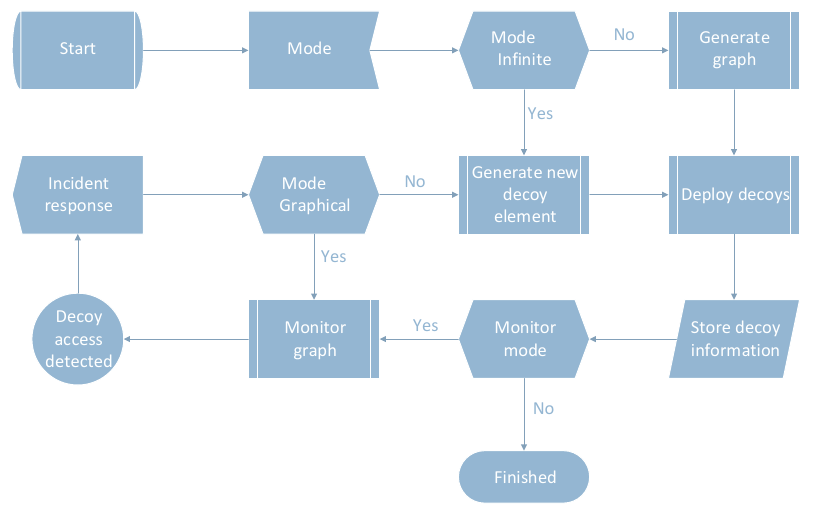}
\end{figure}

\subsection{Selection of Decoy Elements}\label{subsec.selection}
For the reference implementation, the following six decoy types have been selected:
\begin{itemize}
	\item[(a)] Comment ($D_1$): In the reference implementation, the decoy type comment is interpreted as information embedded in a source code file.
	\item[(b)] DB-entry ($D_2$): In this work, DB-entries are stored locally in a \textit{SQLite}. Depending on the application, a full database management system (DBMS) might be more plausible and could be implemented as well.
	\item[(c)] Account ($D_4$): For this decoy type a OS user account is created. A reference corresponding to an account is implemented as a hint for an account name and the password. If no password is provided, the password is set to be the account name. This requires an attacker to try basic password guessing attacks in order to follow the reference. Once authenticated successfully, the subsequent references are implemented implicitly by placing the referenced files inside the home directory of the account. This requires the domain of referenced decoys to be the account. Any transition to a decoy of type $D_4$ results in the creation of a new user account, which will be available as a new domain for subsequently deployed decoys.
	\item[(d)] Document ($D_6$): Document decoys consist of text file with no content other than the subsequent reference itself in the reference implementation.
	\item[(e)] File-metadata ($D_7$): Only the file name itself is used as metadata. The name of a file contains the reference to the subsequent decoy. This type is more complex to observe, as the reference information can be retrieved from the preceding reference and without file access. To mitigate this, the decoy is placed inside a dedicated directory, to which the preceding reference refers. Access to this directory is then monitored.
	\item[(f)] Uniform Resource Locator ($D_8$): This decoy type differs from document and comment type decoys by the type of access. It can be accessed via a web browser and requires to be pointing to a file in the directory of an HTTP server. The provided resource is an \textit{HTML} document, containing a redirect to a non-existing resource. However, the information sources implemented are the \textit{GET}-parameters within the redirect. As described above, these parameters consist of a name and a value, thus being able to provide various decoy references.
\end{itemize}

The transition probabilities set in the reference implementation are given in \cref{tab.transition}. Values are chosen based on the authors’ knowledge of the system and may be subject to adaption and optimization.

\begin{table}[h!t]
	\renewcommand{\arraystretch}{1.5}
	\centering
	\caption{Transition matrix $P$ for a selection of decoy elements}
	\label{tab.transition}
	\begin{tabular}{l l l l l l l l}
		\toprule
		\multicolumn{2}{c}{} & \multicolumn{6}{c}{Potential outbound decoy type}\\
		 & & $D_1$ & $D_2$ & $D_4$ & $D_6$ & $D_7$ & $D_8$ \\
		 \multirow{6}{*}{\rotatebox[origin=c]{90}{Decoy type}} & $D_1$ & 0.15 & 0.15 & 0.1 & 0.25 & 0.15 & 0.2 \\
		  & $D_2$ & 0 & 0.3 & 0.1 & 0.3 & 0 & 0.3 \\
		  & $D_4$ & 0.25 & 0.25 & 0 & 0.25 & 0.25 & 0 \\
		  & $D_6$ & 0.2 & 0.25 & 0.05 & 0.15 & 0.15 & 0.2 \\
		  & $D_7$ & 0 & 0.2 & 0.3 & 0.2 & 0 & 0.3 \\
		  & $D_8$ & 0 & 0 & 0.2 & 0.8 & 0 & 0 \\
		\bottomrule
	\end{tabular}
\end{table}

As a result, a total of 26 possible transition are implemented.

\subsection{Digging the Hole: Decoy Deployment}\label{subsec.digging}
The deployment process can be split up into different phases, which are elaborated on in the following section. After that, the three implemented operational modes are introduced and discussed with visual examples.

\subsubsection{Generation and management}
When new decoy elements need to be deployed, the proposed system will execute four steps: First, based on the decoy element type containing the reference, the subsequent decoy type is determined. The employed probability distribution is shown in \cref{tab.transition}. Additionally, in this step the number of outbound references is determined, if necessary. This is the case in infinite and hybrid mode. Second, the referenced decoy elements are created. Those elements are created as specified in the implementation discussed above. However, they are also stored as an entry in the management database. In this step, these referenced decoy elements are also indicated as subsequent to the containing decoy element in the management database. Third, the actual reference to the previously created decoy elements are generated and deployed into the containing decoy element. Last, the containing reference will be added to the \textit{inotify} watch list. If access to this decoy element is observed, this process is repeated for each subsequent element. The subsequent elements then become decoy elements, containing references themselves. \par
For each deployed and monitored element, the DB-entry contains information if access to the decoy element has been observed, to prevent more than one creation of references for the same subsequent decoys.

\subsubsection{Operation Modes}
\begin{itemize}
	\item[(a)] Graphical: (1) In graphical mode, a generator based on a graphical model is used to create a directed network. Nodes represent decoys and edges represent references from a decoy to another. (2) Then, an initial node is chosen randomly, where the probability $p$ for a node $n \in N$ is $p = \dfrac{1}{|N|}$. Afterwards, the decoy type and the domain are determined for the initial node by a similar procedure. (3) The network is traversed by an iterative deepening depth-first search (IDDFS), where for each transition, the type of the subsequent decoy is determined based on the transition probability matrix P. IDDFS is chosen because it provides a reasonable trade-off between the speed of a depth-first search (DFS) and the completeness of a breadth-first search (BFS). If the networks are small and memory is not the primary concern, BFS may also provide reasonable results. DFS is not suitable, as it does not guarantee to traverse each node within the network. (4) After the search is completed the decoy reference network is ready for deployment. This mode requires the graph to be connected. If the underlying model does not guarantee the graph to be connected, this property can be ensured as proposed earlier. An overview of the graphical mode is given in \cref{fig.graphical}, where the decoy type is represented by the node color.
	\item[(b)] Infinite: (1) In infinite mode, one single decoy is initially deployed and monitored. This decoy contains references to a particular number of subsequent decoys. The referenced decoys are not monitored, and do not contain further references. (2) If an access to the initially deployed and monitored decoy is observed, the decoys referenced by it are assigned to a particular number of subsequent decoys. As in the step before, these referenced decoys do not contain references. However, the decoys referenced by the initial decoy, now contain the references to their subsequent decoys. (3,4) This procedure can be repeated infinite times, generating a deep tree structure of decoys and references. An overview of this process is given in \cref{fig.infinite}, where the decoy type is represented by the node color. \par
	In contrast to the graphical mode, there is only one inbound edge per node. From a conceptional perspective, a new reference can be set to refer to a previous decoy element. In this case, the management database needs to be queried to determine all deployed and monitored decoys. From this set of active decoys a number of decoys can be chosen. The infinite mode ensures connectivity of the iteratively generated graph by design. If non-observable ($\neg P2$) decoy types are used in this mode, they must be considered as accessed and the subsequent decoys have to be deployed. This must be repeated until only decoy types satisfying $P2$ are the latest deployed type within a branch.
	\item[(c)] Graphical-infinite (hybrid): Both presented modes have inherent disadvantages. The graphical mode has a limited number of decoys and those decoys have been deployed without knowledge about the attacker. The infinite mode has only one initial entry node, while the number of decoys is related to the probability of an attacker triggering it. To mitigate these disadvantages, both modes can be combined into the graphical-infinite mode, where the initialization phase is based on the graphical mode and the operational phase (monitor and deploy) is based on the infinite mode. However, only decoy types satisfying $P4$ can be extended by further references. $\neg P4$ types are considered as in graphical mode.
\end{itemize}

\begin{figure*}
	\caption{Overview of graphical mode, (1) generating network, (2) determining initial node,(3) traversing network, (4) completing network}
	\label{fig.graphical}
	\centering
	\includegraphics[scale=1]{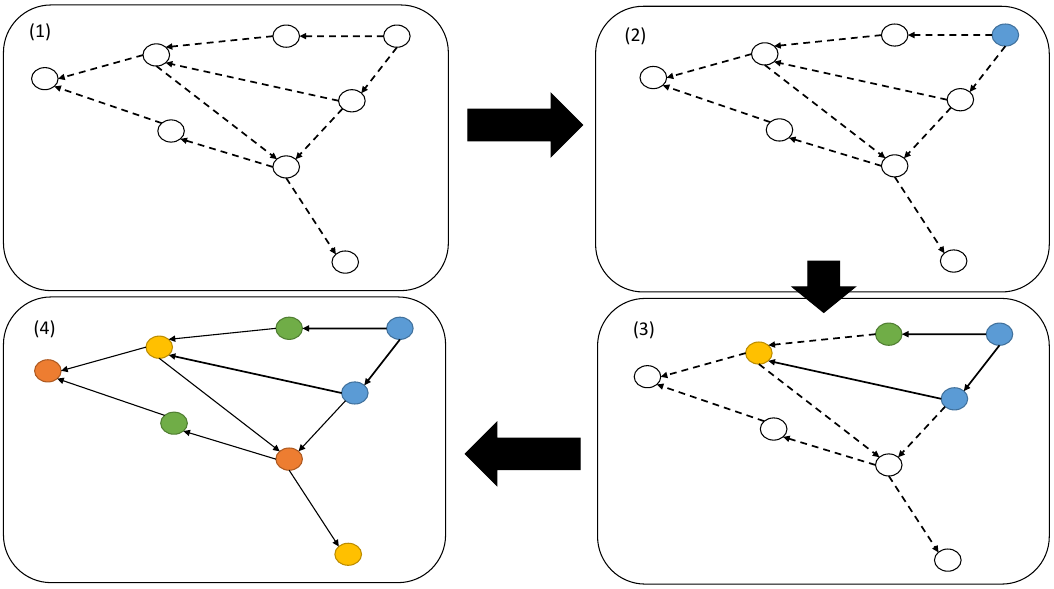}
\end{figure*}

\begin{figure*}
	\caption{Overview of infinite mode, (1) generating initial decoy, (2,3,4) monitoring and deploying decoys}
	\label{fig.infinite}
	\centering
	\includegraphics[scale=1]{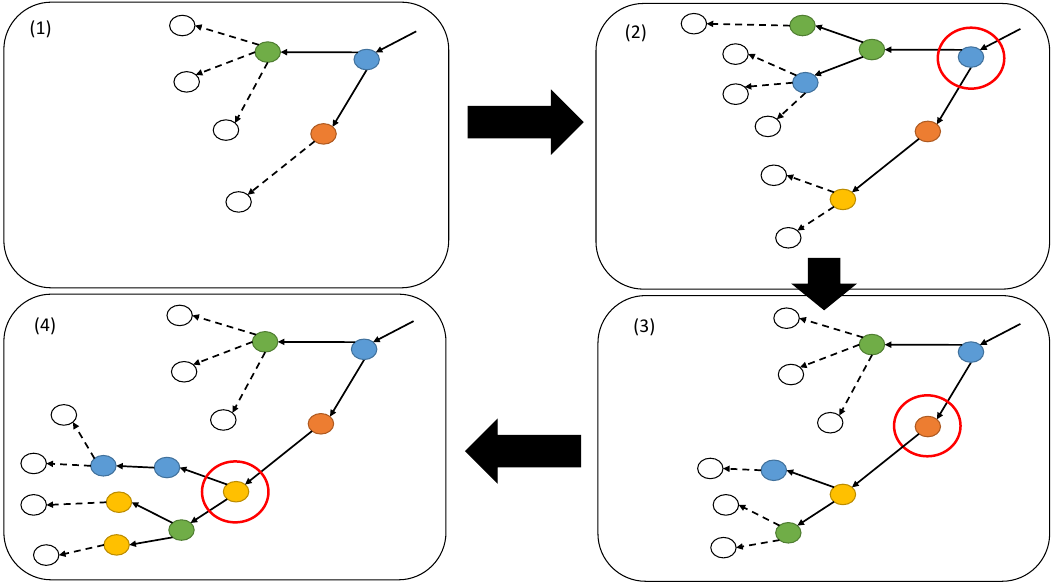}
\end{figure*}

\section{Evaluation and Analysis}\label{sec.evaluation}
To evaluate the proposed system, three different aspects are considered. First, the relationship between attacker and system behavior in the context of the introduced learning mechanism is discussed. Then, security and performance perspectives on the system are elaborated separately. An empirical evaluation of the effectiveness of the concept and it's different aspects is yet to be done. 

\subsection{Behavioral Model of Attackers}\label{subsec.behavioral}
An attacker can be described by establishing a \textit{Markov} model, where $P_A$ is a matrix containing the probabilities of an attacker to follow an existing reference. However, in contrast to the transition matrix $P$ of the decoy network, the probabilities are not based on the plausibility of the reference itself but on the:

\begin{itemize}
	\item[(a)] knowledge an attacker has about computer systems of the type the attacker believes to be interacting with, as this knowledge may limit the capabilities of an attacker to follow a certain reference and the
	\item[(b)] objective an attacker intents to accomplish on the system as this affects the likelihood an attacker follows a reference to a particular decoy type. For example, an attacker looking for network resources is more likely to follow a network-related reference.
\end{itemize}

The implemented learning mode will adapt $P$ to $P_A$ . It will eliminate references an attacker will not follow and enhance frequently followed references. This behavior points out the need for a moderate learning rate $\eta$ to prevent over-fitting. In the conducted experiments 0.3 was found to be sufficient, however, the rate should be adapted to the use case. In general, a higher number of expected observations should result in a lower value of $\eta$. A limitation of this approach is that interaction with a decoy element is not necessarily initiated by an intruder. Although deception systems are characterized by having few to none false-positive results, unintentional interaction cannot be ruled out and therefore an interaction with a decoy element, while being a strong indicator for malicious activity, is not definitive proof for it. However, the low false-positive rate should prevent alarm fatigue.

\subsection{Security Analysis}\label{subsec.security}
As the reference implementation is based on software, it might contain vulnerabilities. In this subsection, the system concept instead of the reference implementation is reviewed from a security perspective. There are two major aspects that are considered in this analysis: Security boundaries and evasion. A security boundary is trespassed if an attacker is able to gain access to information that is supposed to be inaccessible by the attacker. The most plausible attack vector is the monitoring daemon. This daemon is required to be executed in privileged mode and is unable to drop privileges after initialization. Therefore, compromising the daemon results in a fully compromised system. The daemon is only required in the infinite mode or if intrusion detection is necessary. If this is not the case, program execution is finished after the initialization, thus reducing the time frame for a successful attack. \par 
System evasion is possible, as \textit{inotify} is not polling the file system for performance reasons. Thus, accesses and modification invoked by \textit{mmap}, \textit{msync} and \textit{munmap} are invisible to the system. Additionally, pseudo-file systems like \texttt{/proc} are not monitorable. Both restrictions may be circumvented when using a different monitoring system.
Intentional boundary trespassing and evasion require an attacker to suspect the system to employ a deception framework. There are various types of detection techniques to identify deception systems \cite{Bahram.2010, Corey.2004, Corey.2005, Dahbul.2017, Dornseif.2004, Ferrand.2013, Holz.2005, Krawetz.2004, Morris.2018, Sysman.2015}. However, several countermeasures may be applied to conceal the operation.

\begin{itemize}
	\item[(a)] Decoy meta information: Despite being of deceptive nature, file meta information may provide information to an attacker. This can be prevented by determining suitable metadata based on the environment. Context-sensitivity is the objective of previous research work \cite{Fraunholz.2018on, Fraunholz.2017an, Fraunholz.2017towards, Whitham.2017}, and not part of this work.
	\item[(b)] Process information: An attacker with system access may identify suspicious processes running on the system. Countermeasures are closely related to rootkits, trojans and backdooring techniques. DLL-injection, for example, can be applied to hide the process inside another process.
	\item[(c)] System files: Besides the decoy elements itself, the deception framework requires a few files to be used. In the reference implementation, a \textit{Python} script exists on the system as well as a number of templates for the decoy generation and a \textit{SQLite} database to store information about the deployed decoy elements. Obfuscation techniques may be adapted to increase the detection complexity. At the cost of persistence, the required files may only be stored within volatile memory. Even if volatile memory is not resistant against memory forensics, it is more complex to investigate and less likely to raise suspicion.
\end{itemize}

\subsection{Performance Implications}\label{subsec.performance}
From a memory consumption perspective, the framework can be considered light-weight. The source code consists of a few thousand lines of \textit{Python} code, with less than 200,000 bytes in size. The DB used to store and manage active decoys is based on \textit{SQLite}, where the data stored is given in \cref{tab.structure}. \textit{SQLite} does only support particular data types and is therefore subject to unnecessarily large memory consumption. Boolean types are not available and boolean information needs to be stored in 1 byte integer values. However, \textit{SQLite} database file headers contain only 100 bytes of information, thus allowing for fast and memory preserving operation. 

\begin{table*}[h!t]
	\renewcommand{\arraystretch}{1.5}
	\caption{Structure of the data base for decoy management}
	\label{tab.structure}
	\begin{tabular*}{\textwidth}{@{\extracolsep{\stretch{2}}}*{5}{l}@{}}
		\toprule
		Name & Data type & Size & Usage & Example \\
		\cmidrule{1-5}
		Identifier & Int & 1 byte & Primary key & 1 \\
		Created & Int & 1 byte & Monitor state & True \\
		Reference set & Int & 1 byte & Children set & True \\
		Decoy type & Int & 1 byte & File prop. & $D_7$ \\
		Path & Char & 100 byte & Locate decoy & \texttt{/tmp/doc\_384.txt} \\
		Children & Char & 10 byte & Child ID & 2,3 \\
		Domain & Char & 40 byte & File perm. & user1 \\
		Parents & Int & 1 byte & Parent ID & 1 \\
		\bottomrule
	\end{tabular*}
\end{table*}

Storage formats such as \textit{ext3}, \textit{ext4}, \textit{HFS+}, \textit{Brts} and \textit{ZFS} have a maximum size of a file name of 255 byte. However, the path size may extend the file size. \textit{POSIX} requires the maximal path size to be more than 255, but does not give an upper limit. \textit{Linux} for example uses 4096 bytes as maximal supported path size. Determining the size of an occurring path in advance is therefore not suitable. In the reference implementation, 100 bytes are used to store path information. In total, 155 bytes are stored for each deployed decoy. The default page size of data stored by \textit{SQLite} is 4096 bytes, which is enough space for around 26 decoy elements. So the data base will grow by 4096 bytes for approximately each 26th decoy element in the default configuration. As a result, the total memory consumption of the reference implementation in operation is less than 250,000 byte, thus rendering the proposed solution suitable for large-scale deployment as well as deployment in resource restricted environments. \textit{Inotify} consumes system resources as well. Most important are the watch list and the event queue. The watch list contains the inode and the mask. It is growing as the number of files to monitor is increasing. By default, the maximal number of watches is limited to 8, 192 watches per instance, while 128 instances per user are allowed. The event queue is populated based on the watch list. Events exceeding the maximal number of events are dropped. This number is defined to be 16,384 by default. In experiments conducted during this work, \textit{inotify} was able to monitor 100,000 files on \textit{COTS}-hardware without a significant reduction of performance.

\section{Conclusion}\label{sec.conclusion}
In this work, the concept and deployment of interreferencing decoy elements was introduced. The concept was developed as a fast deployable, environment agnostic stand-alone system, whereat the reference implementation was created for UNIX environments. To the best of the authors’ knowledge, an interreferential decoy network has not been addressed in previous research. An advantage of the proposed solutions are an increase in interaction between an attacker and the deployed decoy elements, increasing the probability of inducing the attacker uncertainty, while wasting the attacker's time and resources. A disadvantage is the decrease in quality of the observations as interaction and therefore the threat intelligence are affected by the references. The proposed system is able to enhance the intrusion detection process as well as delaying and interfering ongoing reconnaissance activities. It will be extended to include honeypots as endpoint nodes in the interreferential network. For operational usage a \textit{SIEM}-conform \textit{API} will also be provided.
\section*{Acknowledgment}
This research was supported by the German Federal Ministry of Education and Research (BMBF) within the SCRATCh project under grant number 01IS18062E. The SCRATCh project is part of the ITEA 3 cluster of the European research program EUREKA. The responsibility for this publication lies with the authors.

\bibliographystyle{IEEEtran}
\bibliography{rabbit-hole}

\end{document}